\documentclass[preprint,authoryear]{elsarticle}
\usepackage[margin=2.5cm]{geometry} 

\usepackage[utf8]{inputenc}
\usepackage{amsmath,amssymb}
\usepackage{graphicx}
\usepackage{setspace}
\usepackage{array}
\usepackage{multirow}
\usepackage{hyperref}
\usepackage{rotating}
\usepackage{subcaption}
\usepackage{adjustbox}
\usepackage{threeparttable}
\usepackage{booktabs}
\usepackage{siunitx}
\usepackage{graphicx}
\usepackage{amssymb}
\usepackage{setspace}
\usepackage[utf8]{inputenc}
\usepackage{array}
\usepackage{natbib}
\usepackage{amsmath}
\usepackage{multirow}
\usepackage{hyperref}
\usepackage{rotating}
\usepackage{subcaption}
\usepackage{adjustbox}
\usepackage{threeparttable}
\usepackage{comment}

\usepackage{booktabs}

\usepackage{siunitx}

\usepackage{tabularx}
\usepackage{booktabs} 
\usepackage{array}    

\usepackage{color}
\definecolor{redcolor}{rgb}{1.0,0.,0.}

\definecolor{bluecolor}{rgb}{0,0.,1}

\begin{document}


\title{Auditing automated research assessment: an interpretable machine learning approach to validate funding criteria}



\author[usp]{Rafael P. Gouveia}
\author[ucb]{Thiago C. Silva}
\author[usp]{Diego R. Amancio}

\address[usp]{Institute of Mathematics and Computer Science, Universidade de S\~ao Paulo, S\~ao Carlos, Brazil}
\address[ucb]{Universidade Católica de Brasília, Brasília, Distrito Federal, Brazil}

\newpage 


\begin{abstract}
This paper empirically examines the practical validity of the official evaluation criteria underpinning the Research Productivity (PQ) Grant framework, as governed by the Brazilian National Council for Scientific and Technological Development (CNPq). By operationalizing regulatory dimensions (including bibliographic output, human resource training, and scientific recognition) as measurable variables extracted from CVs and OpenAlex bibliometric data, we treat policy-defined indicators as testable hypotheses rather than a priori assumptions. Using a block-based adaptation of the Boruta feature selection algorithm across several machine learning classifiers, we evaluate the statistical contribution of each dimension in distinguishing grant levels, with a focus on identifying top-tier (Level 1A) researchers. Our models achieve high predictive performance, with mean AUC scores reaching 0.96, indicating that PQ levels carry a robust and structured statistical signal. However, explanatory power is heavily concentrated within a limited subset of features, specifically bibliographic production, graduate-level supervision and institutional management roles. Conversely, several criteria explicitly emphasized in the regulations demonstrated no detectable statistical contribution to classification outcomes. These findings reveal a potential misalignment between the formal regulatory framework and the effective signals driving evaluation outcomes, suggesting that the practical evaluative signal is substantially more compact than officially stated and providing evidence-based insights for the refinement and transparency of research assessment policies.
\end{abstract}

\maketitle








\frenchspacing



\doublespacing

\section{Introduction}

Research funding agencies rely on structured evaluation frameworks to identify and reward scientific productivity, yet the criteria underpinning these frameworks are rarely subjected to systematic empirical scrutiny~\citep{leiden_manifesto}. In Brazil, the Research Productivity Grant (PQ) awarded by the National Council for Scientific and Technological Development (CNPq) plays a central role in shaping academic careers, influencing access to funding, prestige and institutional visibility. The grant is governed by detailed regulations that specify multiple dimensions of academic activity (such as bibliographic production, human resources formation and scientific recognition) as well as time windows and career-stage constraints that purportedly guide evaluation decisions. 

Although PQ levels are widely used as indicators of scientific excellence, it remains unclear to what extent the formally prescribed criteria align with the effective signals observed in evaluation outcomes. In practice, criteria may differ in their empirical relevance due to redundancy, measurement limitations, or correlations across dimensions and timeframes. As a result, the operational logic by which \emph{productive researchers} are identified may deviate from the intent expressed in official regulations, with significant implications for transparency, equity, and the interpretability of evaluation decisions.

Existing literature has extensively mapped the profiles of grant holders across diverse disciplines, including public health, engineering, physiology, etc, typically employing bibliometric and demographic data to characterize the distributions of awards. However, these investigations often treat the official criteria as background information rather than variables to be tested, leaving the actual mechanics of the PQ evaluation process still unexplained~\citep{saude_publica, fisiologia, enfermagem, geografia, farmacologia, engenharia, outro_lattes}. While scientific success is frequently predicted using parsimonious models based on h-indices or publication counts, few studies actually test the practical validity of the grant levels themselves within a policy-defined framework. This work addresses this gap by empirically examining the relevance of regulation-defined evaluation criteria in the context of PQ assessments within Computer Science. 

Our study seeks to answer the following research questions: {first, to what extent can PQ grant levels be accurately recovered as a structured statistical signal from curriculum and bibliometric data? Second, are there mandated dimensions of productivity that appear redundant or weakly informative in practice? And finally, do evaluation standards remain uniform across different decision thresholds, such as the distinction between baseline productivity and the identification of top-tier 1A researchers? }

The methodology of our paper involves operationalizing the officially stated criteria as measurable variables and treating their practical impact on evaluation outcomes as hypotheses to be tested via machine learning. To this end, features derived from these criteria are extracted from structured curriculum vitae data and complementary bibliometric sources such as OpenAlex. A structured feature selection strategy is then applied to these features, utilizing a block-based adaptation of the Boruta algorithm to assess the statistical contribution of each criterion and its distinct temporal representations, while accounting for correlated predictors and repeated observations.
Our results showed that most grant decisions can be predicted using curriculum and bibliometric data and that, though there are some differences between the tasks of discerning baseline productivity or top-tier productivity, for both of them, extremely reduced feature sets can be obtained that contain all the signals which practical usefulness can be detected.

By focusing on the empirical signal carried by different dimensions of productivity, this study provides insight into which aspects of academic activity effectively differentiate PQ recipients and which appear redundant within observed evaluation outcomes. This approach goes beyond just characterizing researcher profiles, contributing to broader discussions on the design and assessment of research productivity indicators and the evidence-based refinement of academic reward systems. Ultimately, the findings demonstrate that the practical evaluative signal may be substantially more compact than the formal framework implies, offering a data-driven path toward greater transparency and efficiency in research assessment.

\section{Related Works}

Several studies have explored the feasibility of predicting scientific success (typically quantified through bibliometric indices and awards) using author-level metrics. For instance, \cite{einstein} demonstrated substantial predictive power ($R^2 = 0.67$) using a parsimonious linear model to forecast neuroscientists' h-indices five years into the future. Their model relied on a limited set of features, including the current h-index, career age (time since first publication), total article count, journal diversity, and the number of publications in high-impact journals. However, the study observed variations in performance across different researcher cohorts, suggesting that success criteria may be discipline-specific. 
In contrast, \cite{fortunato} examined the cross-disciplinary application of cumulative metrics, such as the h-index, as predictive targets. They concluded that the predictive power of these features is frequently overestimated due to their inherent autocorrelation, rendering them suboptimal for identifying the fundamental drivers of productivity and success. Furthermore, the study found that these models often rely on career age as a proxy for more substantial predictors, which degrades performance when evaluating high-achieving early-career scientists. This highlights the necessity of disentangling the complex interactions between variables in this type of research.

National and regional indicators are frequently employed by researches as a metric to evaluate academic performance success. A notable example of one of those indicators is Brazil’s CNPq Research Productivity (PQ) Grant, a prestigious award used to recognize scientific excellence and research impact. Existing literature has mapped the profiles of grant holders across diverse disciplines, including public health, physiology, nursing, geography, pharmacology, and engineering~\citep{saude_publica, fisiologia, enfermagem, geografia, farmacologia, engenharia, outro_lattes}. Typically, these investigations employ bibliometric (publication counts, citation metrics and h-indices), demographic, geographical and/or institutional data to compare distributions across grant levels and regions, depending on the goals of the research. Together, these studies offer critical insights into the structural dynamics and inherent inequalities of the Brazilian research landscape~\citep{geociencia}. 
Evidence consistently indicates a high concentration of grants in the Brazilian Southeast, suggesting potential disparities in the funding distribution process~\citep{fisiologia, enfermagem, geografia, farmacologia, geociencia}. Furthermore, bibliometric metrics (such as citation counts and publication volume) are consistently and positively correlated with grant levels across the disciplines evaluated \citep{saude_publica, farmacologia, geociencia, engenharia, outro_lattes}. In contrast, metrics related to student supervision and demographic attributes, such as gender, do not show such a consistent relationship, with results varying significantly across studies.

While PQ status is widely used to measure productivity, few studies actually test the validity of the grant levels themselves. Instead, research typically describes who gets which level based on demographics or citations, treating the official criteria as background information rather than variables to be tested~\citep{saude_publica, enfermagem, farmacologia}. This leaves the actual mechanics of the PQ evaluation process largely unexplained.

Methodological approaches to studying the relationship between research productivity and PQ awards are often rooted in the analysis of structured curriculum vitae data, specifically via the Lattes Platform. As a standardized repository of Brazilian academic careers, the platform allows researchers to extract key indicators such as bibliographic output, postgraduate supervision, and research funding. These metrics are frequently augmented with citation data from external databases like OpenCitations, Google Scholar, Scopus, or Web of Science~\citep{outro_lattes}. However, due to the inherent difficulties in processing qualitative criteria and narrative sections, most studies prioritize quantifiable features to ensure methodological reproducibility.

The temporal modeling of scientific productivity represents a significant theme within the literature. Existing studies frequently use fixed time windows (typically three or five years) reflecting the emphasis placed on recent output by funding and promotion committees. These models often adjust for career stage or the time elapsed since the completion of a doctorate~\citep{outro_lattes}. Regarding PQ grants, while official regulations explicitly differentiate temporal horizons based on grant level—prioritizing longer trajectories for senior categories—empirical research often adopts these prescriptions a priori. Consequently, most analyses collapse productivity into a single window or assume the evaluative relevance of prescribed timeframes without subjecting them to direct empirical testing.

More recently, the assessment of research has increasingly integrated machine learning and data science techniques to model academic productivity, collaboration networks, and grant-related outcomes~\citep{tohalino2022predicting,brito2023analyzing}. 
For instance, \cite{coautoria} examined co-authorship structures among PQ grant holders, while \cite{altmetric} explored the utility of altmetrics as alternative impact indicators. Although these computational approaches provide more granular representations of scholarly activity, they often prioritize descriptive mapping or predictive accuracy over a critical interrogation of the formal evaluation criteria themselves. While feature selection and interpretability techniques are well-established within the machine learning domain, their application as tools for auditing the policy-defined frameworks of academic reward systems remains underdeveloped.

In contrast to previous research focused on description or prediction, this study evaluates the PQ framework itself. By translating official evaluation criteria into measurable features and applying structured feature selection, we identify which dimensions of academic productivity provide a genuine statistical signal and which are redundant for predicting PQ awards. Upon integrating regulation-based data extraction, temporal modeling, and interpretable machine learning, this work provides empirical evidence on how scientific productivity is practically measured within the Brazilian funding system. 

\section{Materials and methods}

Our methodology comprises four main steps, which are detailed in the following sections and summarized below.

\begin{enumerate}

\item \emph{Dataset}: this section describes the dataset, which comprises PQ grant levels alongside bibliographic information and data extracted from the authors' \textit{curricula vitae}.

\item \emph{Feature extraction}: this section details the processes of feature extraction and manipulation. To this end, we construct multiple views of the data to capture information within specific time windows of interest. Additionally, bibliographic data are processed to compute annual metrics (such as the $h$-index) and to generate co-authorship graphs, which facilitate the extraction of specialized collaboration features.

\item \emph{Feature selection}: a feature selection process is employed using machine learning to identify the features that provide the most significant predictive signal for classification.

\item \emph{Machine learning methods}: this section describes the six machine learning models employed in the feature selection phase -- comprising four ensemble methods and two baseline classifiers -- alongside specific details of the training process. 

\end{enumerate}

\subsection{Dataset}

We developed a dataset that combines Research Productivity grant results with the specific metrics defined in official evaluation rules. For each researcher, we extracted the relevant performance features (such as publication counts and supervision history) within the exact timeframes used by the funding agency to make award decisions. The researcher productivity data were primary extracted from the CV Lattes platform~\citep{lattes}.  

To identify grant recipients along with their respective approval years and fellowship levels, we consulted the CNPq Open Data repository~{\citep{dados_abertos}} in July 2025. This allowed for the compilation of a comprehensive list of all PQ grants approved between 2002 and 2023. Subsequently, relevant features were extracted from the Lattes platform ({specifically using the dataset scraped by~\cite{dataset_lattes}}) by matching names, institutions, and research areas. To enable a more granular analysis of the grant's regulatory framework, the final dataset was narrowed to include only researchers in the field of Computer Science. 

Finally, leveraging the publication records extracted from the Lattes platform, we mapped each author to their corresponding profile on OpenAlex~\citep{OpenAlex}, which is one of the largest citation networks currently available. This enabled the extraction of advanced bibliographic metrics, including longitudinal citation data and the relative impact of the venues where each researcher published their work.

{\subsection{Feature extraction}}

The feature extraction process was designed to strictly adhere to the official regulations for the Productivity (PQ) grant~\citep{edital}. In the standard review process, evaluators first examine a summary of qualifications highlighting a researcher’s most significant achievements; if deemed viable, the complete curriculum vitae (from the Lattes platform) is then analyzed. Due to the lack of a structured dataset for the initial summary -- and assuming its primary role is workload reduction -- this study ignores the first step. Consequently, the majority of features were extracted directly from the Lattes platform, supplemented by citation and impact data from OpenAlex. Features absent from the structured fields of the Lattes CV, such as those found only in free-text sections like project roles and social impact were excluded from the dataset unless they could be retrieved from specialized bibliometric platforms. {Tables \ref{Table1} through \ref{Table5} in the Supplementary Information (SI) provide a comprehensive breakdown of the features specified in the regulations for each category versus those extracted from the data. Asterisks denote features requiring data from OpenAlex, while dashes indicate those that were excluded. The following sections provide an overview of the proposed features and the methodology adopted for their collection. 

{The specific criteria for the Research Productivity Grant in Computer Science fall into five broad categories: \emph{bibliographic production}, \emph{human resource training}, \emph{national and international recognition}, \emph{technological, social, and governmental contributions}, and \emph{participation and funding of research projects/networks}. \emph{Bibliographic production} include indicators of publication volume, venue type (journal \textit{vs} conference and foreign \textit{vs} domestic) and venue quality i(f it is among the top 10\% venues referenced in the dataset in mean value of citation per article in a period of two years), as well as metrics indicating production regularity (average yearly publications), intellectual leadership (number of publications as first-author), research topic (proportion of non–computer-science publications) and scientific impact. 

Scientific impact was captured through citation-based metrics, including total citations, author-level indices (h-index and i10-index), and the number of field-normalized highly cited publications (being among the top 10\% globally), while collaboration patterns were summarized through measures of repeated co-authorship and the diversity of collaborators and institutions. \emph{Human-resource training}  included counts of student supervisions at different levels, participation in examination committees, course and instructional material development, and involvement in editorial boards, peer review, advisory committees, and scientific event organization. Additional dimensions of \emph{national and international recognition}  included awards, management roles, and service to funding agencies. Finally, \emph{technological social and governmental} contributions were captured through technologies, patents, and software registrations (domestic and international), while \emph{participation and funding of research projects/networks} were represented by number of funded research projects participated and collaboration indicators derived from co-authorship networks.}

The regulations specify evaluation periods of 5 years for junior-level grants and 10 years for senior-level grants, with earlier records used primarily for tie-breaking or exceptions. Accordingly, values for all relevant features previously mentioned were collected for each of these three time frames (totaling three features by attribute). Additionally, because certain restrictions apply based on the time elapsed since the conferral of the doctorate,the time since the researcher's first doctorate was also obtained via their CV. 

The grant tier for each researcher was retrieved annually from the CNPq (National Council for Scientific and Technological Development) official open-access repository. These grants are organized into a hierarchy: Level 2 represents the entry-level tier, while Level 1 is subdivided into 1D, 1C, 1B, and finally 1A, the latter being the most prestigious rank. Typically, researchers within any Level 1 subcategory are regarded as senior investigators with established leadership in their respective fields. Hereafter, they are referred to collectively as Level 1X researchers.

\ 

\subsection{Feature selection}
\label{section:feature_selection}

To identify the features truly relevant to researcher classification, we use a variation of the Boruta feature selection algorithm~\citep{boruta}. This method was chosen because the goal of feature selection in the experiment is not merely to improve predictive performance, but to assess the practical validity of specific indicators defined by the PQ grant regulations. Methods such as LASSO, recursive feature elimination, or standard tree-based importance typically perform minimal-optimal selection, meaning they aim to find the smallest subset of variables that preserves predictive accuracy. When features are correlated, as is common with indicators measured over different time windows, these methods may arbitrarily retain one variable while discarding others that carry essentially the same signal. This makes them poorly suited for evaluating whether a regulated metric itself has empirical relevance. The Boruta framework is preferable because it performs \emph{all-relevant feature selection}, meaning that each variable (or, in the case of out custom implementation, block of variables) is statistically tested against a noise baseline to determine whether it contains information about the outcome beyond random fluctuations. As a result, Boruta does not attempt to compress the feature set, but instead identifies all indicators that demonstrably carry signal, even when they are redundant with others.

The core mechanism relies on creating \emph{shadow features} through the random permutation of original values across observations {(e.g., in Figure \ref{fig:shadow}, the shadow version of feature 1 is constructed using block A from entry 7 and block B from entry 2)}.
This permutation severs any structural relationship between a feature and the classification outcome, effectively establishing a robust noise baseline. This approach is ideal because it produces irrelevant yet realistic variables that preserve the original range, variance, sparsity, and empirical distribution. Maintaining these characteristics is essential, as importance measures in ensemble models like Random Forests are highly sensitive to distributional traits. Using purely synthetic noise, such as Gaussian-distributed values, would introduce bias, discrepancies in scale, skewness, or cardinality could otherwise artificially distort the resulting importance scores.

\begin{figure}[hbtp!]
    \centering
    \includegraphics[width=1.1\linewidth]{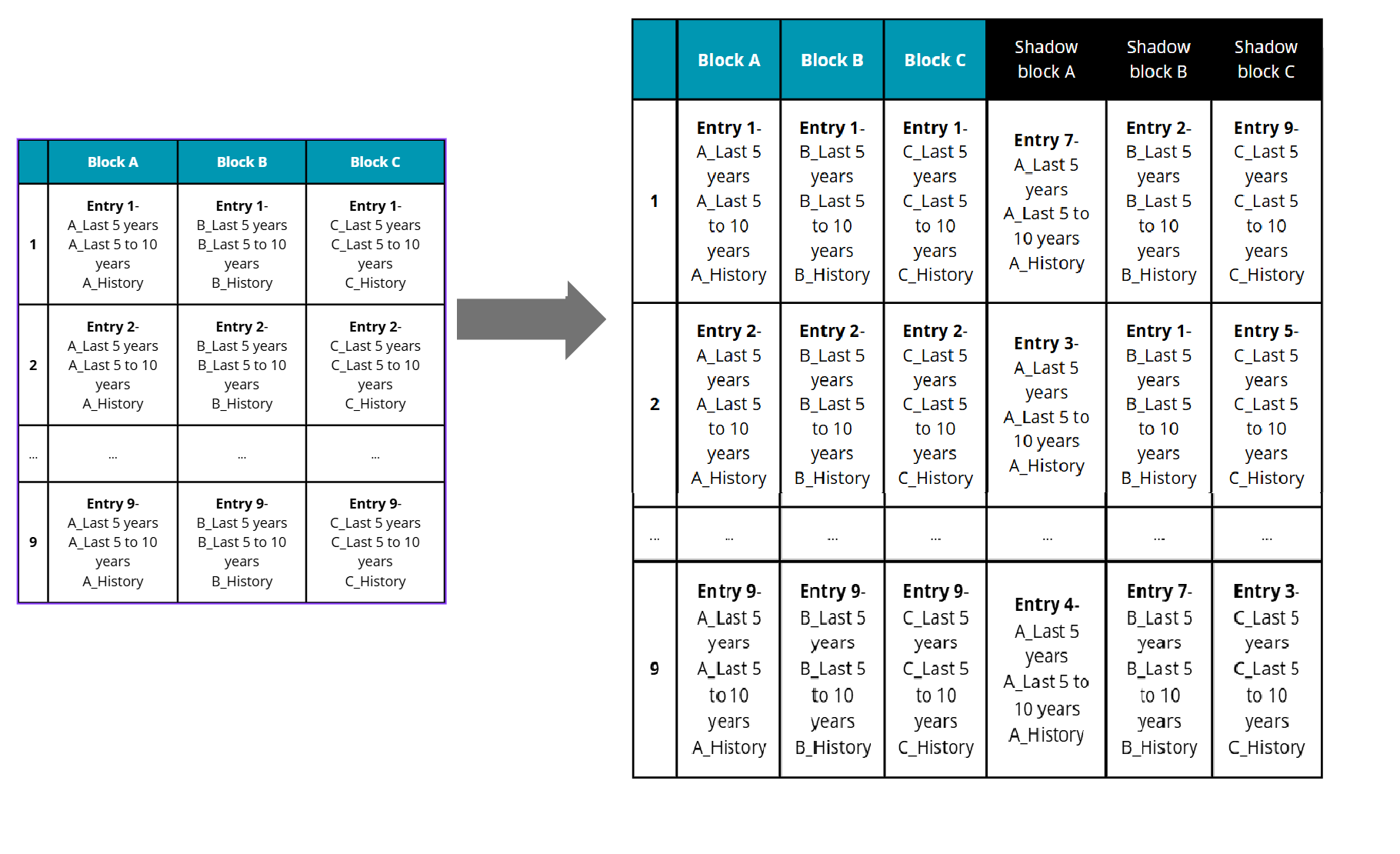}
    \caption{Boruta's process of creating shadow features by permuting the features (or, in this case, blocks of features) across observations to create realistic noise.}
    \label{fig:shadow}
\end{figure}

The algorithm employs an iterative process: shadow features are generated, importance scores are calculated, and statistical tests determine which features significantly outperform the most important shadow feature. Features confirmed as significant are retained, those significantly underperforming are rejected, and the remainder are re-evaluated in the subsequent iteration.

Our implementation follows the process illustrated in Figure \ref{fig:iterate}. Across multiple iterations of shadow feature generation and model training, we calculate the permutation importance for each feature, which is defined as the reduction in classifier performance when a variable's structural information is disrupted via row-wise shuffling (Figure \ref{fig:a}). For each original feature, we then count how often across iterations its permutation importance exceeds that of the most important shadow feature in that round; such instances are recorded as 'hits' (Figure \ref{fig:b}). Finally, as shown in Figure \ref{fig:c}, the hit rates are evaluated using two-tailed binomial tests to determine if they differ significantly from 0.5. We apply the Benjamini-Hochberg procedure to adjust for the false discovery rate across multiple feature blocks.
If a feature block's null hypothesis (that it is better than than the best noise only half the time, which indicates that it itself is merely noise) is rejected, the feature block is considered a significant feature if the sample hit-rate is bigger than 0.5 and it is considered an insignificant feature (and is removed from the feature set) if the hit-rate is lower than 0.5. If the null hypothesis is not rejected, the feature block is deemed tentative and kept at the model for the next iteration alongside the features deemed significant. After a pre-determined set of iterations all tentative blocks are deemed insignificant and the process ends}. 

\begin{figure}[htbp]
    \centering

    \begin{subfigure}{\linewidth}
        \centering
        \includegraphics[width=0.9\linewidth]{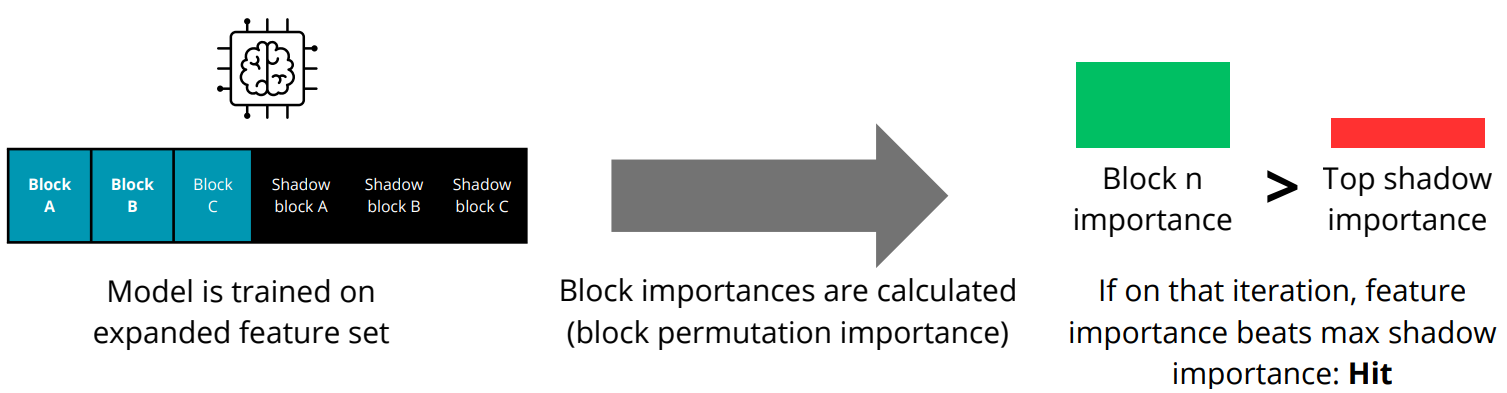}
        \caption{Comparison to noise baseline}
        \label{fig:a}
    \end{subfigure}

    \vspace{0.5cm}

    \begin{subfigure}{\linewidth}
        \centering
        \includegraphics[width=0.45\linewidth]{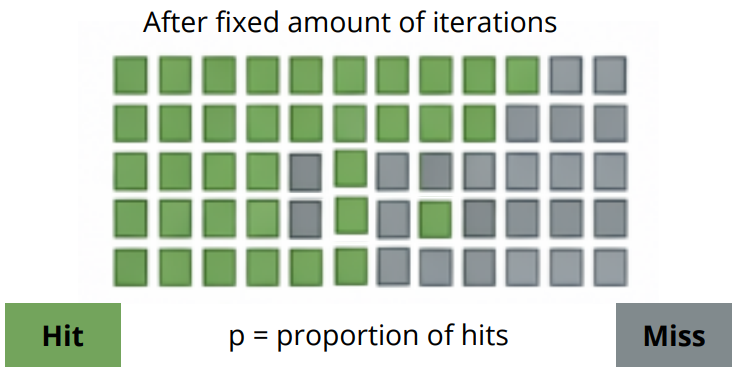}
        \caption{Sample gathering over different dataset folds and row permutations}
        \label{fig:b}
    \end{subfigure}

    \vspace{0.5cm}

    \begin{subfigure}{\linewidth}
        \centering
        \includegraphics[width=0.66\linewidth]{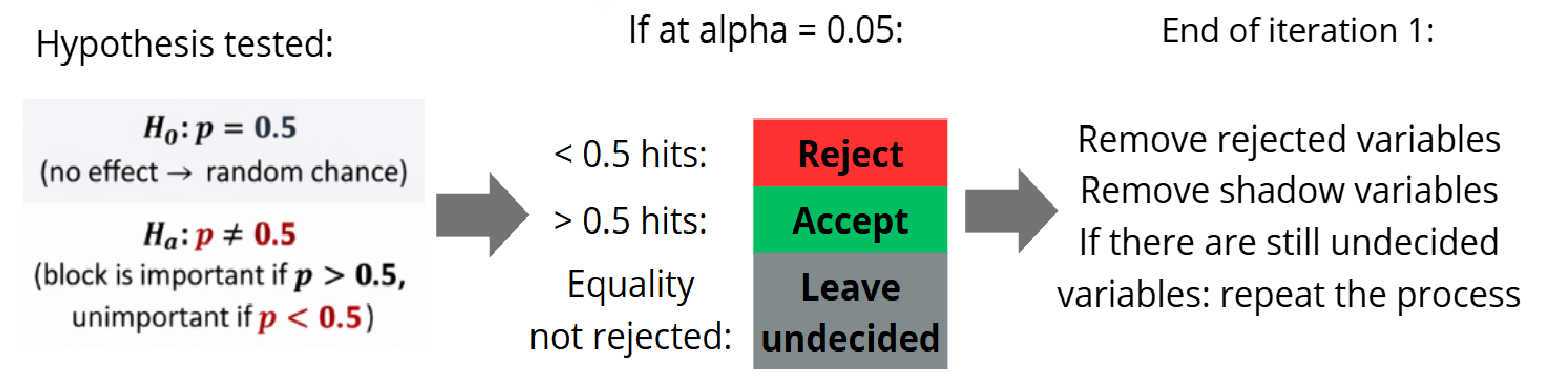}
        \caption{Hypothesis testing for rejection or acceptance of features to the next iteration}
        \label{fig:c}
    \end{subfigure}

    \caption{Boruta's cycle of pruning features by comparing them to noise. The entire Figure set refers to one single overall Boruta iteration}
    \label{fig:iterate}
\end{figure}


{Our custom implementation seeks to enhances the standard Boruta approach by grouping highly correlated features into cohesive blocks and, during the whole process, the features in the block are treated as one, being shuffled at the same time, both for the creation of shadow features and for the permutation importance calculation.}
{This modification addresses a primary limitation of the traditional implementations of the algorithm based on variations of permutation importance: the distortion of importance in the presence of multicollinearity among the features. In the standard method, features are permuted individually, one at a time. 
A limitation of this estimation method is that highly correlated predictors can mask a feature's true importance. If the model can infer a feature's information from other preserved variables, its importance score may not drop significantly. Consequently, a feature might be incorrectly excluded because it offers no marginal utility relative to the existing set, rather than being excluded only if it lacks intrinsic predictive power~\citep{random_forest}.
Furthermore, \cite{regions} argues that permutation-based methods, by breaking dependence among correlated features, generate points from sparse and/or unrealistic regions of the feature space and give undue importance to them, which forces the model to extrapolate to regions with little to no data, causing distortions. By grouping these features and destroying them together, we ensure they are only removed if no predictive signal is detected across the entire block, resulting in a more robust and conservative pruning process.}

To generate the feature blocks for the selection process, we employ a two-stage grouping process. First, we identify \emph{fundamentally connected} features to form initial candidate groupings (specifically, those theoretically expected to be correlated or mutually predictive). In this study, these consist of features representing the same attributes across the three distinct timeframes: the most recent five years, five to ten years ago, and more than ten years ago (e.g., publication counts).
Second, we calculate the Spearman Correlation between these feature blocks, defined as the maximum correlation between any of all the possible pairs of features across blocks. This design choice favors an aggressive grouping strategy, ensuring a more conservative pruning of feature blocks during the subsequent selection phase.

After determining the correlations between the preliminary blocks, we merge them into final correlated groups. This process is framed as a weighted clique partitioning problem. In simple terms, the objective is to organize the features into distinct clusters (or "cliques") such that the items within each cluster are as highly correlated to each other as possible. To ensure these groups are meaningful, we impose two conditions: 
\begin{enumerate}

\item \emph{Maximize similarity}: we want the sum of correlations within the groups to be as high as possible.

\item \emph{Minimum threshold}: no two features can be placed in the same group if their correlation is lower than a set limit, ensuring that every member of a group is genuinely related to every other member.

\end{enumerate}
Formally, let $S \in \mathbb{R}^{n \times n}$ be a matrix where each element $S_{ij}$ represents the absolute correlation between preliminary blocks $i$ and $j$. We define $z_{ij} \in \{0, 1\}$ as a binary decision variable where $z_{ij} = 1$ if blocks $i$ and $j$ are assigned to the same group, and $z_{ij} = 0$ otherwise. Given a correlation threshold $\tau$ (set to $0.6$ for this experiment), the optimization problem is expressed as:

\begin{align}
\text{maximize} \quad 
& \sum_{i<j} S_{ij} z_{ij}  \label{problem}
\\[6pt]
\text{subject to} \quad 
& z_{ij} = 0
&& \forall i<j \text{ such that } S_{ij} < \tau
\\[6pt]
& z_{ij} + z_{ik} - 1 \le z_{jk}
&& \forall i<j<k \label{t1}
\\
& z_{ij} + z_{jk} - 1 \le z_{ik}
&& \forall i<j<k \label{t2}
\\
& z_{ik} + z_{jk} - 1 \le z_{ij}
&& \forall i<j<k \label{t3}
\\[6pt]
& z_{ij} \in \{0,1\}
&& \forall i<j. \label{domain}
\end{align}
Equations \eqref{t1}, \eqref{t2}, and \eqref{t3} represent the transitivity constraints. These ensure that if blocks $A$ and $B$ are grouped together, block $C$ is included in that group if and only if it is also grouped with both $A$ and $B$, thereby maintaining consistent partitions. As hierarchical clustering is not explored in this study, these preliminary groupings are flattened into disjoint clusters. These clusters serve as the fundamental predictive units in our block-based Boruta implementation, where they are evaluated, potentially pruned, and permuted as singular entities.

\ 

\subsection{Machine learning methods}

To ensure that the feature relevance identified by the Boruta procedure was independent of any specific learning algorithm, the analysis was replicated using four tree-based ensemble classifiers: eXtreme Gradient Boosting (XGBClassifier)~\citep{xgboost}, Random Forest (RandomForestClassifier)~\citep{random_forest}, Light Gradient Boosting Machine (LGBMClassifier)~\citep{lgbm}, and Extra Trees (ExtraTreesClassifier)~\citep{extratrees}. Tree ensembles were selected because the original Boruta implementation relies on Random Forests ~\citep{boruta}, and these models naturally provide stable feature-importance estimates in scenarios characterized by high redundancy and non-linear interactions ~\citep{random_forest}. Consequently, the inability of such models to detect a signal serves as a strong indicator of its actual absence, a conclusion that is more difficult to draw with less robust models. For comparison, two linear classifiers were also evaluated: a Support Vector Machine (LinearSVC) \citep{svm} and a Logistic Regression model \citep{scikit}. Achieving a strong fit and consistent feature selection with these simpler models would further validate the hypothesis of a stable underlying signal. 

{The evaluated models were configured using hyperparameters with an emphasis on stability rather than extensive tuning. For the tree-based ensembles, both Random Forest and the Extra Trees classifier were implemented with 500 trees, unrestricted depth, and feature subsampling set to the square root of the number of variables, ensuring robust and stable importance estimates. The gradient boosting models, namely XGBoost and LightGBM, were configured with 300 boosting rounds, a learning rate of 0.05, and both row and column subsampling set to 0.80, balancing predictive performance and generalization while controlling overfitting. Finally, for the linear models, the Logistic Regression was used with standard $L_2$ regularization ($C = 1.0$) and so was the linear Support Vector Machine. Across all models, random seeds were fixed to ensure reproducibility where applicable.}

The model training process accounted for statistical dependencies arising from multiple observations of the same researcher over different time windows. To prevent information leakage, cross-validation was performed using grouped splits based on the researcher’s Lattes platform ID. 
Specifically, a grouped k-fold strategy ensured that all observations from a single researcher were assigned to the same fold during both initial and final performance estimation. This grouped approach was also applied within the Boruta process where necessary. 
By evaluating models on researchers not seen during training, we avoided artificially inflated performance and generated more reliable feature-importance estimates.

\subsection{Experimental setup}

Our experimental design evaluates feature relevance across two tasks: distinguishing researchers exceeding baseline productivity (non-Level 2) and identifying top-tier (Level 1A) individuals. The goal is to identify possible discrepancies between features required by regulations and those that are statistically significant in practice. To maintain the validity of our results, we employ a grouped sampling strategy; all records for an individual researcher are kept within the same fold to avoid data contamination.

The feature selection is replicated across various classifiers to ensure the robustness of the identified subset. Following this, we test for model degradation using the reduced feature set. The study concludes with an exploratory analysis of high-productivity profiles, using dimensionality reduction and clustering to investigate whether multiple, distinct archetypes of productivity coexist in the data.

\section{Results and discussion}

{\subsection{Models performance and feature correlation analysis}
\label{results_start}
Following data collection and feature extraction, the first step was to evaluate classifier performance through meaningful baselines. This evaluation serves to validate the initial feature set and provides a benchmark for assessing potential performance degradation during subsequent feature pruning. We employed various tree ensemble models (consistent with the Boruta algorithm) alongside linear models for validation. Table \ref{tab:auc_comparison1} presents the classification results for the distinction between Level 2 and Level 1 (1A, 1B, 1C, and 1D) researchers. Additionally, it details the model's performance in discriminating Level 1A researchers from all other award-holding categories.

\begin{table}[htbp]
    \caption{Classification performance (mean AUC $\pm$ standard deviation) across 5-fold cross-validation for the Level 1 vs. Level 2 and Level 1A vs. remaining classification tasks, utilizing the complete feature set. Note: '1X' denotes all Level 1 award-holding categories.}
    \centering
    \label{tab:auc_comparison1}
    \small
    \begin{tabular}{lcc||cc}
        \hline
        & \multicolumn{2}{c||}{1X vs 2} & \multicolumn{2}{c}{1A vs remaining} \\
        \cline{2-5}
        Estimator & Mean AUC & Std AUC & Mean AUC & Std AUC \\
        \hline
        XGBClassifier        & 0.9062 & 0.0249 & 0.9656 & 0.0201 \\
        LGBMClassifier       & 0.9015 & 0.0277 & 0.9613 & 0.0185 \\
        RandomForestClassifier & 0.8913 & 0.0301 & 0.9412 & 0.0267 \\
        ExtraTreesClassifier & 0.8901 & 0.0273 & 0.9452 & 0.0307 \\
        LinearSVC            & 0.8354 & 0.0277 & 0.8758 & 0.0282 \\
        LogisticRegression   & 0.8081 & 0.0191 & 0.8898 & 0.0290 \\
        \hline
    \end{tabular}
\end{table}

Notably, the classifiers demonstrate high performance, which enables a meaningful interpretation of the feature set and the individual impact of specific variables. It is also worth noting that distinguishing researchers whose productivity is only slightly above the grant baseline (1X vs.~2) is significantly more challenging than identifying top-tier researchers (1A vs.~remaining). Overall, these results suggest that the selection of awardees could be largely automated, particularly for the identification of Level 1A grantees.

To initiate the experiment, it was necessary to identify and group correlated features. As described in Section \ref{section:feature_selection}, features representing different time slices of the same attribute were categorized as \emph{naturally related}. We then computed the max-pooled Spearman correlation between these groups, with the result being illustrated in the heatmap in Figure \ref{fig:heatmap}. The full heatmap can be seen in the supplementary material (\autoref{fig:heatmap_full}).


\begin{figure}[htbp]
    \centering
    \includegraphics[width=1.0\linewidth]{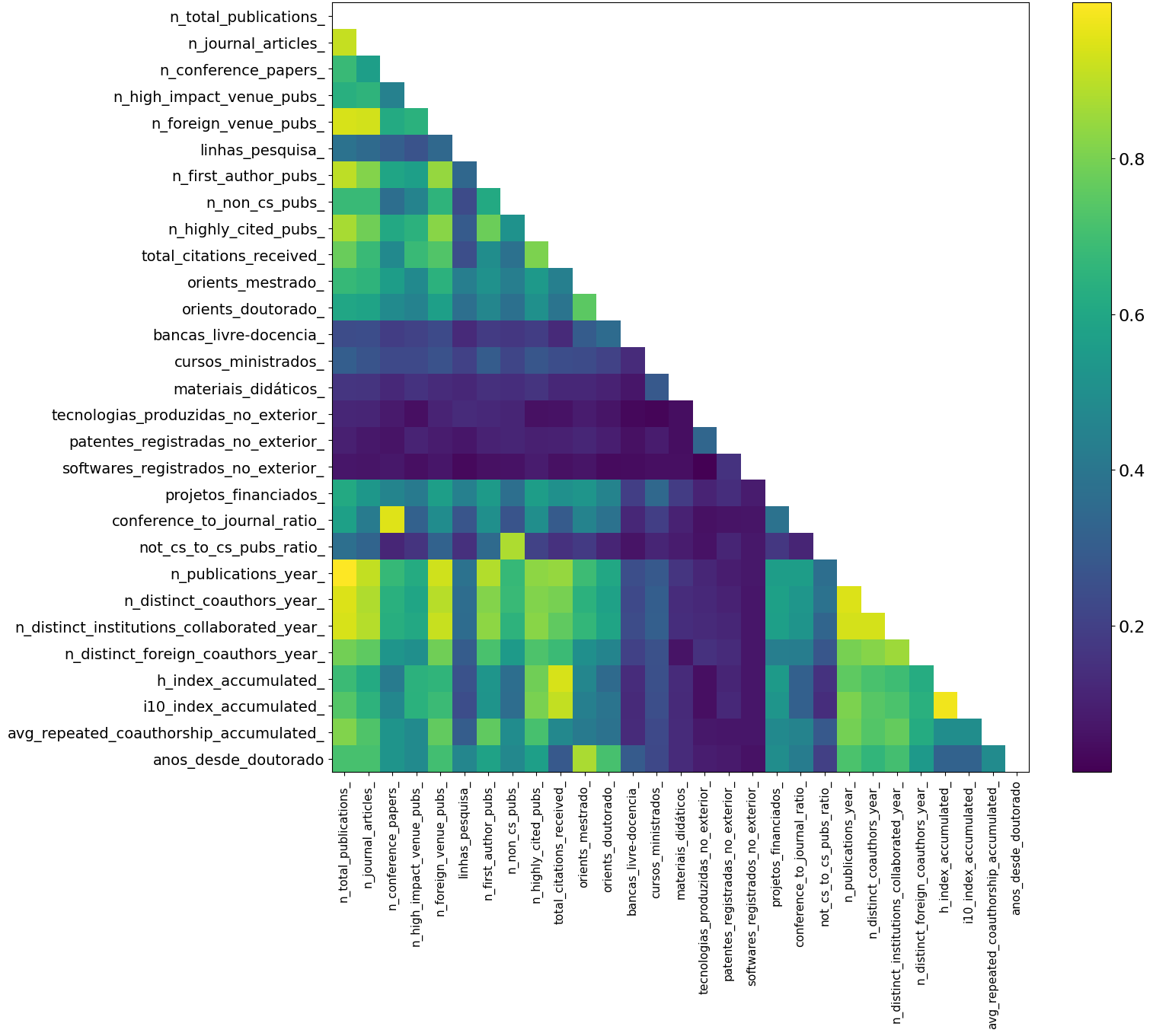}
    \caption{Heatmap of maximum absolute correlations between different features across time windows.}
    \label{fig:heatmap}
\end{figure}

Most bibliographic metrics (regardless of their specific nature) and those pertaining to collaboration networks exhibit strong mutual correlations. This robust interdependence justifies their aggregation into cohesive units, a process subsequently formalized through the linear programming optimization. These results further suggest that while individual metrics possess distinct nuances, researchers who excel in one dimension generally perform well across others, indicating a centralized signal of academic productivity. In contrast, technological production features demonstrate an unexpectedly weak correlation with all other variables except among themselves. This isolation suggests these features are largely independent of traditional productivity measures. Such independence may be inherent to the nature of technological output or, alternatively, it could be a spurious artifact of underreporting within the Lattes CVs. If the latter is true, it highlights a need for increased awareness regarding the strategic importance of documenting technological contributions. A similar, albeit less pronounced, pattern was observed for special courses, habilitation examination boards, and instructional materials. Given the nature of these activities, this lack of correlation likely stems from systematic underreporting. Finally, the effective age of the researcher (years since first doctorate) shows an extremely high correlation with most variables. The strong association with Master's supervisions is particularly noteworthy, suggesting that career age may serve as an effective proxy for overall cumulative productivity within the current evaluation framework.

The correlation between variables are further elucidated through optimal feature set merging. By solving the weighted clique partitioning problem (equations \ref{problem} through \ref{domain}) based on the max-pooled correlations, we obtained the groupings shown in Table \ref{tab:feature_groups}. 
In the resulting clusters, most bibliographic metrics grouped together, with notable exceptions for metrics related to conference publications, research topics, and leadership roles (i.e., whether the researcher leads projects or merely participates). These three categories formed distinct groups, consistent with their specific design objectives. It is worth noting, however, that general bibliographic metrics also included metrics relating to collaborative research networks, which by itself shows that productivity and research networks have intrinsic correlations, despite being organized as different categories. This outcome retroactively justifies the decision to adjust the feature selection process to account for feature correlations, since otherwise, many of these variables might have been spuriously rejected. Furthermore, it was unexpected to find that career length (defined as the time since the first doctorate) grouped with master's supervisions and participation in common postgraduate examination boards.
A possible explanation for this association is that these metrics capture the steady-state phase of an academic career. In the Brazilian context, progression from Level 2 to Level 1 may involve the substantial accumulation of “service” indicators (such as master’s thesis defenses and committee participation) which naturally correlate with time since the PhD. The fact that these features cluster together suggests that they share similar variance, making them highly effective for distinguishing between Level 1 (1X) and Level 2 researchers, but less discriminative for identifying the exceptional bibliometric outliers characteristic of the Level 1A group.

\begin{table}[htbp]
\centering
\caption{Features grouped by correlation according to the solution of the Spearman correlation-weighted clique partitioning problem described in Section \ref{section:feature_selection}. ``Other individual features'' represent those features that were not clustered together.}
\label{tab:feature_groups}
\begin{tabular}{ll}
\hline
\textbf{Group theme} & \textbf{Features} \\
\hline

\multirow{5}{*}{Bibliographic metrics} & total publications, journal publications, foreign venue publications,\\
   & highly cited publications, total citations received, \\
   & average yearly number of distinct coauthors, average yearly publications,\\
   & average yearly number of distinct institutions collaborated,\\
   & average yearly number of distinct foreign coauthors, h-index, i10 index; \\
\hline
Conference metrics & conference publications, conference to journal publication ratio; \\
\hline

\hline
Leadership metrics & publications as first author, average repeated co-authorship degree; \\
\hline
\multirow{2}{*}{Topic metrics} & non computer science publications,\\
   & computer science to not computer science publication ratio; \\
\hline
\multirow{2}{*}{Graduate human resources} & master's supervisions, master's examination boards,\\
   & doctorate examination boards, years since first doctorate; \\
\hline
Events organization & national events organization, international events organization; \\
\hline
\multirow{9}{*}{Other individual features} & high impact venue publications; lines of research; doctorate supervisions; \\
 & postdoctoral supervisions; undergraduate research supervisions; \\
 & habilitation examination boards; special courses; editorial boards; \\
 & instructional materials; journal peer review bodies; advisory committee; \\
 & commission board; awards and honors; management positions;  \\
 &  peer reviewing for funding agencies;  technologies made abroad; \\
 & technologies made in Brazil; patents registered abroad;\\
 & patents registered in Brazil; softwares registered abroad;  \\
 & softwares registered in Brazil; funded projects. \\
\hline
\end{tabular}
\end{table}

The feature selection algorithm detailed in {Section \ref{section:feature_selection} was applied to the four} tree ensemble models from the initial benchmark (Table \ref{tab:auc_comparison1}). The results are summarized in Table \ref{tab:feature_freq_comparison}. For clarity, feature groups with multiple members are referred to by their \emph{group theme}, as defined in Table \ref{tab:feature_groups}.
While some disagreement between models was expected due to their differing inductive biases (which motivated the use of multiple models rather than relying solely on the Random Forest implementation of Boruta~\citep{boruta}), the four models showed substantial consensus regarding both feature acceptance and rejection. This was particularly evident in the more straightforward task of detecting the top-productivity researchers among all. Notably, across both tasks, bibliographic metrics, graduate human resources, and management positions were unanimously accepted. 

\begin{table}[htbp]
    \caption{Feature selection frequency across classifiers for the 1X vs. 2 and 1A vs. remaining scenarios. Notably, while several features are significant in both classification settings, others -- such as editorial board membership and leadership metrics -- are only relevant for discriminating between Levels 1 and 2. }
    \label{tab:feature_freq_comparison}
    \scriptsize
    \begin{tabular}{lccccc||ccccc}
        \hline
        & \multicolumn{5}{c||}{1X vs 2} & \multicolumn{5}{c}{1A vs Remaining} \\
        \cline{2-11}
        Feature Group 
        & RF & XGB & LGBM & ET & Selection (\%) 
        & RF & XGB & LGBM & ET & Selection (\%) \\
        \hline
        
        \textbf{Bibliographic metrics}
        & \textbf{$\checkmark$} & \textbf{$\checkmark$} & \textbf{$\checkmark$} & \textbf{$\checkmark$} & \textbf{100}
        & \textbf{$\checkmark$} & \textbf{$\checkmark$} & \textbf{$\checkmark$} & \textbf{$\checkmark$} & \textbf{100} \\

        \textbf{Graduate human resources}
        & \textbf{$\checkmark$} & \textbf{$\checkmark$} & \textbf{$\checkmark$} & \textbf{$\checkmark$} & \textbf{100}
        & \textbf{$\checkmark$} & \textbf{$\checkmark$} & \textbf{$\checkmark$} & \textbf{$\checkmark$} & \textbf{100} \\

        Undergraduate Research Supervisions 
        & $\checkmark$ &  &  & $\checkmark$ & 50 
        & \textbf{$\checkmark$} & \textbf{$\checkmark$} & \textbf{$\checkmark$} & \textbf{$\checkmark$} & \textbf{100} \\

        Journal Peer Review Bodies 
        & $\checkmark$ & $\checkmark$ & $\checkmark$ &  & 75 
        & \textbf{$\checkmark$} & \textbf{$\checkmark$} & \textbf{$\checkmark$} & \textbf{$\checkmark$} & \textbf{100} \\

        Advisory Comittee 
        &  &  &  & $\checkmark$ & 25 
        & \textbf{$\checkmark$} & \textbf{$\checkmark$} & \textbf{$\checkmark$} & \textbf{$\checkmark$} & \textbf{100} \\

        \textbf{Management Positions}
        & \textbf{$\checkmark$} & \textbf{$\checkmark$} & \textbf{$\checkmark$} & \textbf{$\checkmark$} & \textbf{100}
        & \textbf{$\checkmark$} & \textbf{$\checkmark$} & \textbf{$\checkmark$} & \textbf{$\checkmark$} & \textbf{100} \\

        Doctorate Supervisions 
        & \textbf{$\checkmark$} & \textbf{$\checkmark$} & \textbf{$\checkmark$} & \textbf{$\checkmark$} & \textbf{100}
        & $\checkmark$ &  &  & $\checkmark$ & 50 \\

        Lines Of Research 
        &  & $\checkmark$ & $\checkmark$ &  & 50 
        &  & $\checkmark$ & $\checkmark$ &  & 50 \\

        Awards And Honors 
        &  &  &  &  & 0 
        &  & $\checkmark$ & $\checkmark$ &  & 50 \\

        Conference metrics 
        &  & $\checkmark$ &  &  & 25 
        &  &  &  &  & 0 \\

        High Impact Venue Publications 
        &  &  &  &  & 0 
        &  &  &  &  & 0 \\

        Topic metrics 
        & \textbf{$\checkmark$} & \textbf{$\checkmark$} & \textbf{$\checkmark$} & \textbf{$\checkmark$} & \textbf{100}
        &  &  &  &  & 0 \\

        Special Courses 
        &  &  &  &  & 0 
        &  &  &  &  & 0 \\

        Habilitation Examination Boards 
        & \textbf{$\checkmark$} & \textbf{$\checkmark$} & \textbf{$\checkmark$} & \textbf{$\checkmark$} & \textbf{100}
        &  &  &  &  & 0 \\

        Postdoctoral Supervisions 
        & $\checkmark$ &  &  & $\checkmark$ & 50 
        &  &  &  &  & 0 \\

        \textbf{Leadership metrics}
        & \textbf{$\checkmark$} & \textbf{$\checkmark$} & \textbf{$\checkmark$} & \textbf{$\checkmark$} & \textbf{100}
        &  &  &  &  & 0 \\

        \textbf{Editorial Boards}
        & \textbf{$\checkmark$} & \textbf{$\checkmark$} & \textbf{$\checkmark$} & \textbf{$\checkmark$} & \textbf{100}
        &  &  &  &  & 0 \\

        \textbf{Events organization}
        & \textbf{$\checkmark$} & \textbf{$\checkmark$} & \textbf{$\checkmark$} & \textbf{$\checkmark$} & \textbf{100}
        &  &  &  &  & 0 \\

        Instructional Materials 
        &  &  &  &  & 0 
        &  &  &  &  & 0 \\

        Commission Board 
        &  &  &  &  & 0 
        &  &  &  &  & 0 \\

        Peer Reviewing For Funding Agencies 
        &  &  &  &  & 0 
        &  &  &  &  & 0 \\

        Technologies Made Abroad 
        &  &  &  &  & 0 
        &  &  &  &  & 0 \\

        Technologies Made In Brazil 
        &  &  &  &  & 0 
        &  &  &  &  & 0 \\

        Patents Registered Abroad 
        &  &  &  &  & 0 
        &  &  &  &  & 0 \\

        Patents Registered In Brazil 
        &  &  &  &  & 0 
        &  &  &  &  & 0 \\

        Softwares Registered Abroad 
        &  &  &  &  & 0 
        &  &  &  &  & 0 \\

        Softwares Registered In Brazil 
        &  &  &  &  & 0 
        &  &  &  &  & 0 \\

        Funded Projects 
        & \textbf{$\checkmark$} & \textbf{$\checkmark$} & \textbf{$\checkmark$} & \textbf{$\checkmark$} & \textbf{100}
        &  &  &  &  & 0 \\

        \hline
    \end{tabular}
\end{table}

This subset of three feature blocks carries an unambiguous signal for classification and includes single representatives from three (arguably four) of the five broad regulatory categories: \emph{bibliographic production}, \emph{human resources formation}, and \emph{national and international recognition}. The \emph{bibliographic metrics} block is particularly comprehensive, as it encompasses both primary bibliographic output and collaborative research network data -- two dimensions found to be highly correlated during the preliminary feature grouping phase. Consequently, while research networks are officially situated within the `participation and funding of research projects and networks' category, their statistical alignment with bibliographic metrics leaves `funding' as the sole distinctive attribute of this final regulatory axis.

The convergence of each regulatory axis into a single, unambiguously accepted feature block not only validates the practical relevance of these categories but also suggests that the complexity of an academic trajectory can be synthesized into a parsimonious set of key indicators. From a scientometric perspective, this selectivity indicates that certain metadata serve as robust proxies for latent dimensions of scientific merit. By adopting a less restrictive majority-vote criterion (thereby capturing nuances that individual models might overlook) the set of accepted features expands to include early-career human resources (undergraduate mentorship), editorial board service and the diversification of research lines. This expansion points toward a more multidimensional view of academic impact, encompassing scientific management and the long-term sustainability of the research ecosystem. Conversely, the persistent statistical uncertainty surrounding technological, social, and governmental contributions reveals a critical hiatus: either these dimensions are undervalued, or current data in CVs fail to capture extra-academic impact, leaving the reward system heavily anchored in traditional scientific metrics.

{Table \ref{tab:feature_freq_comparison} also demonstrates that several features were unanimously rejected for both tasks, including \emph{instructional materials} and \emph{special courses}, service on \emph{commission boards} and \emph{funding agencies}, and \emph{technological production in general}. It is of note that the entire axis of technological, social and governmental contributions is completely excluded, which could suggest a complete lack of practical application of consideration of industry-related productivity when considering candidates for the grant, although it is possible that the information missing, such as ``company creation'' and ``demonstrated social impact'' (see Table \ref{Table4}), would be the ones to, in practice, signal the productivity on this area. 
Since these achievements are expressed only within the free-text sections of the \textit{curriculum vitae}, this highlights a limitation of policy automation based strictly on structured data, a common approach to avoid the subjectivity and complexity associated with using LLMs to process unstructured text. Another possibility previously raised is that these features are underreported in Lattes CVs, which complicates classification. Nevertheless, the failure to detect a meaningful signal in domestic or international technology registrations suggests a underemphasis on this productivity axis, which warrants further investigation.
}

It is also noteworthy that the number of \emph{high-impact venue publications} (as measured by two-year mean citedness), which is the proxy for venue quality, was among the feature blocks for which all ensembles failed to capture additional meaningful signal in either task. This surprisingly suggests that venue quality, at least when quantified via \emph{bibliographic metrics} (typically considered a more unbiased metric than qualitative rankings), offers no further discernible impact on classification. This implies that general performance, as measured by the broader \emph{bibliographic metrics} block (e.g., number of publications and citations), remains the primary indicator of bibliographic productivity. Alternatively, it may suggest that venue quality is evaluated using criteria significantly different from bibliographic measures such as mean citedness. 

If the rejection criteria are once again relaxed to a simple majority vote, \emph{conference metrics} are similarly rejected in both tasks. This finding is surprising given the explicit regulatory restrictions concerning the proportion of journal versus conference publications. However, the rejection of these features may be explained by the fact that researchers who have been awarded grants at any level must necessarily satisfy these baseline requirements. Consequently, it can be inferred that once these minimum thresholds are met, the ratio of conference publications exerts no further impact on perceived productivity, suggesting they may carry a weight comparable to that of journal publications.

We also observed that some features provide signal for one task but not the other. This divergence indicates that different weights are applied when distinguishing baseline performance from top-tier productivity. Specifically, the former task shows greater scrutiny toward `inflated' publication counts, as per grant regulations. This is supported by the exclusive detection of signal for \emph{topic metrics} -- which flag out-of-domain works -- and \emph{leadership metrics}, which identify high-volume production in groups where the individual may be making peripheral contributions rather than a lead role. 

In distinguishing baseline productivity, the models also account for a researcher’s capacity to secure funding (a key component of a productive academic profile according to grant regulations) as evidenced by the inclusion of \emph{funded projects}. Furthermore, a broader range of productivity indicators was identified, including \emph{event organization}, \emph{editorial board membership} and \emph{postgraduate supervision}. Of particular interest is the inclusion of \emph{habilitation examination boards}, because despite showing minimal correlation with all other features, this indicator still provides a unique and relevant signal to the classification.

These specific features -- namely \emph{topic metrics}, \emph{leadership metrics}, \emph{funded projects}, \emph{event organization}, \emph{editorial board membership}, and \emph{postgraduate supervision} -- do not influence the classification of top-tier researchers (Level 1A). This occurs despite these metrics constituting a significant portion of the criteria theoretically applicable to all award categories.} {Whether this indicates a blind spot in the evaluation process, where it is incorrectly assumed that candidates for higher tiers do not exhibit the patterns these criteria were designed to detect, remains a compelling topic for future investigation.} {In practice, top-tier classification appears to rely almost exclusively on the features previously identified as common to both tasks, with the addition of \emph{awards and honors} and work on \emph{advisory committees}, which show no significant effect in the "1X vs. 2" (productivity above baseline) scenario. This may be a qualitative property of the dataset, where prestigious awards and jobs are reported only by top-tier researchers, while less significant accolades are either ignored in baseline assessments or have their ``merit'' already encoded within other features.}

\subsection{Performance analysis with reduced feature sets}

To assess the sufficiency of the identified features, the models were retrained using only the reduced feature sets obtained through model consensus for each of the two tasks. This step serves as a validation of the feature selection process, ensuring that the consensus attributes capture the essential discriminative information without compromising predictive power. The performance of these streamlined models was then compared against the original results, as summarized in Table \ref{tab:auc-comparison2}.

\begin{table}[htbp]
    \centering
    \caption{Classification performance (mean AUC and standard deviation over 5
folds) for both 1X vs 2 and 1A vs remaining scenarios using reduced feature sets.}
    \label{tab:auc-comparison2}
    \begin{tabular}{lcc|cc}
        \hline
        & \multicolumn{2}{c|}{1X vs 2} 
        & \multicolumn{2}{c}{1A vs remaining} \\
        \cline{2-5}
        Estimator & Mean AUC & Std AUC 
                  & Mean AUC & Std AUC \\
        \hline
        
        LGBMClassifier      & 0.9083 & 0.0279 & 0.9654 & 0.0131 \\
        XGBClassifier       & 0.9017 & 0.0272 & 0.9676 & 0.0134 \\
        RandomForestClassifier & 0.8933 & 0.0258 & 0.9517 & 0.0222 \\
        ExtraTreesClassifier   & 0.8927 & 0.0263 & 0.9566 & 0.0259 \\
        LinearSVC           & 0.8193 & 0.0233 & 0.9262 & 0.0232 \\
        LogisticRegression  & 0.8016 & 0.0291 & 0.9108 & 0.0129 \\
        
        \hline
    \end{tabular}
\end{table}

A comparison of Tables \ref{tab:auc_comparison1} and \ref{tab:auc-comparison2} reveals no significant performance degradation, despite the removal of most features traditionally mandated by grant regulations. Surprisingly, a performance increase was observed for the linear models in the "1X vs 2" (discerning productivity above baseline) task. To investigate this, a one-tailed paired Student's t-test was conducted; the null hypothesis of equal means was rejected for the Linear Support Vector Machine ($p = 0.00494$), though not for Logistic Regression ($p = 0.0777$) at $\alpha = 0.05$. Given that the performance differences for the other models were even less pronounced, the failure to reject the null hypothesis in those cases is considered trivial. 

The results demonstrate that the reduced feature set selected through the proposed methodology maintains consistent performance, despite the removal of an entire axis of criteria (specifically \emph{technological, social, and governmental contributions}) and numerous other features. This suggests that the excluded attributes do not contribute additional predictive signal. Consequently, it can be concluded that explanatory power is effectively concentrated within a limited subset of features, warranting further investigation into whether this phenomenon arises from limitations in the evaluation process or reflects a natural consequence of the distribution of productivity among researchers.

\begin{figure}[ht]
    \centering
    
    \begin{subfigure}{0.45\linewidth}
        \centering
        \includegraphics[width=\linewidth]{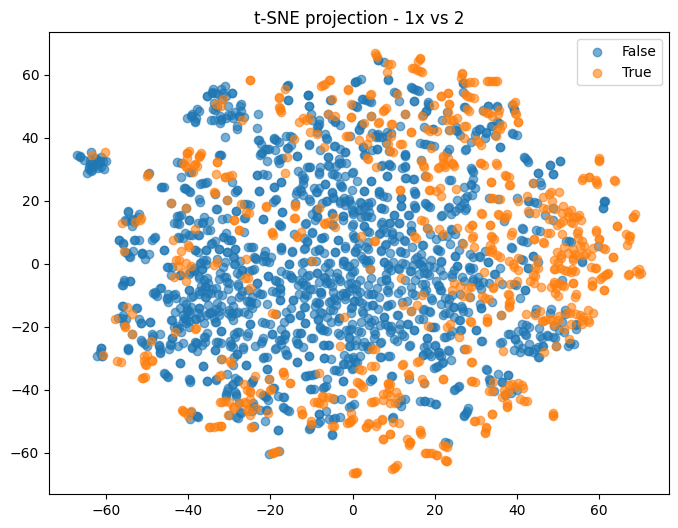}
    \end{subfigure}
    \hfill
    \begin{subfigure}{0.45\linewidth}
        \centering
        \includegraphics[width=\linewidth]{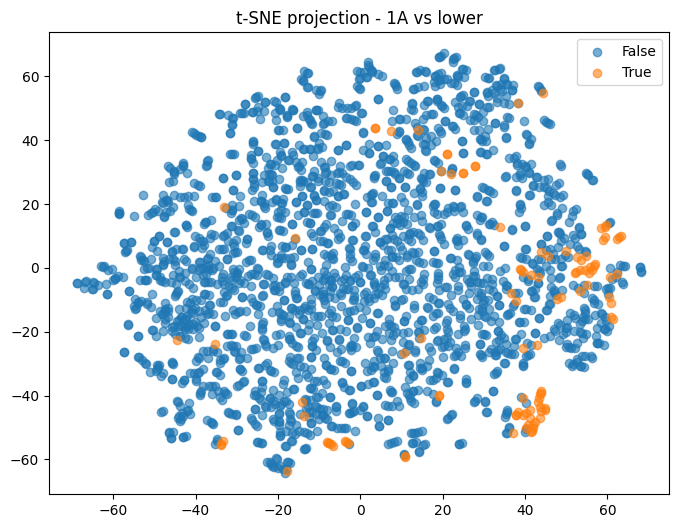}
    \end{subfigure}
    
    \vspace{0.5cm}
    
    \begin{subfigure}{0.45\linewidth}
        \centering
        \includegraphics[width=\linewidth]{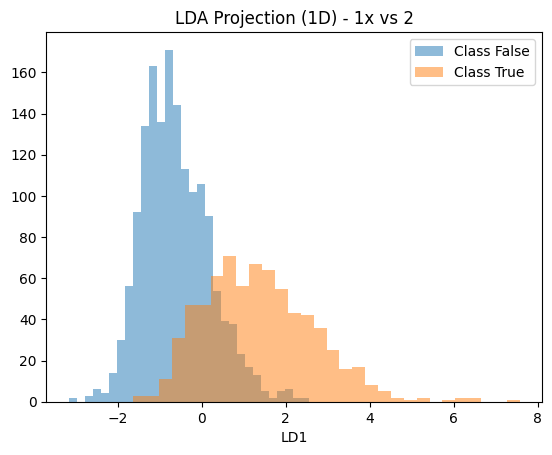}
    \end{subfigure}
    \hfill
    \begin{subfigure}{0.45\linewidth}
        \centering
        \includegraphics[width=\linewidth]{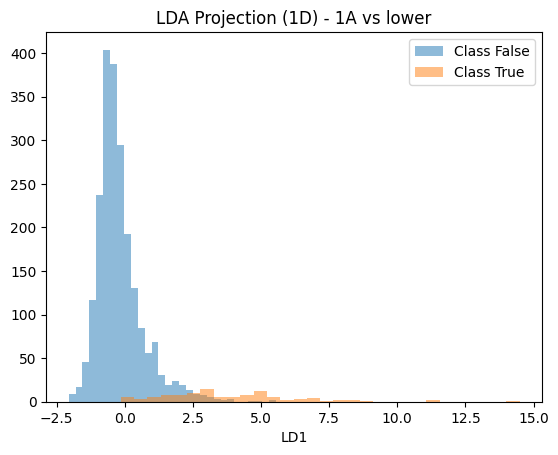}
    \end{subfigure}
    
    \caption{Class separation visualizations for both classification problems using 2D t-SNE and 1D LDA}
    \label{fig:four_images}
\end{figure}

In order to visualize the new feature set, 2D t-SNE and 1D LDA projections are presented in Figure \ref{fig:four_images}. Given the strong performance of linear classifiers, these visualizations illustrate the spatial distribution and separability of the classes for both tasks. The projections reveal a clear disparity in difficulty: while the 1x vs 2 problem shows more discernible class boundaries, the 1A vs lower task exhibits significant overlap and a pronounced class imbalance. Despite these challenges, the feasibility of classification is evident even in these low-dimensional spaces. Notably, the absence of distinct sub-clusters within the classes suggests that productive researchers do not form unique, identifiable 'profiles' based on these features; instead, they represent a relatively homogeneous group in this feature space.

\begin{figure}[htbp]
    \centering
    \includegraphics[width=0.75\linewidth]{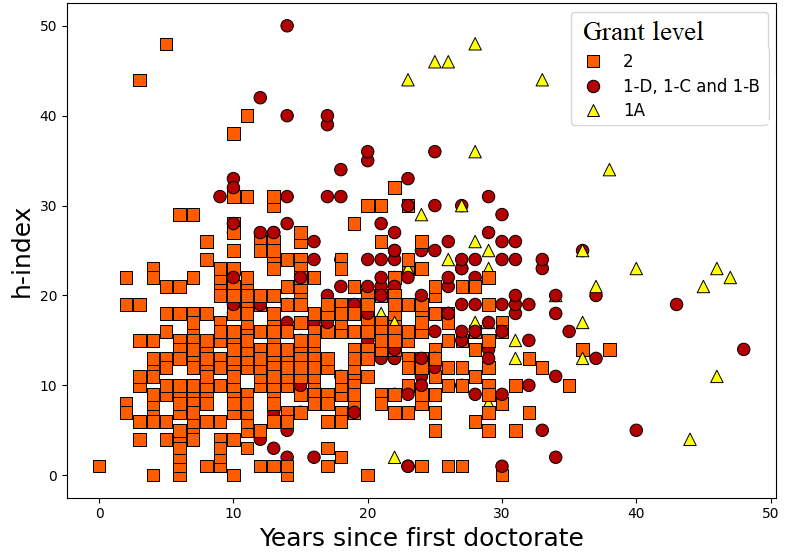}
    \caption{Scatter Plot of years since first doctorate vs h-index colored and shaped by grant level, using the latest data per researcher.}
    \label{fig:scatter}
\end{figure}

As noted in Section \ref{results_start}, the number of years since the doctorate correlates significantly with most features, particularly those later identified as representative. Consequently, investigating this metric as a potential proxy for productivity is a logical progression. Given the prevalence of bibliographic indicators such as the h-index for measuring productivity (the utility of which was validated by our analysis) a scatter plot (Figure \ref{fig:scatter}) was generated. This plot maps the distributions of both variables and their relationship to grant levels to further evaluate their interplay. 
A pronounced correlation is observed between career seniority (quantified as years since the conferral of the doctorate) and grant levels. Level 1A fellowships are predominantly concentrated in the later stages of academic careers, whereas Level 2 grants are clustered toward the earlier years. This distribution is sufficiently structured to allow for the definition of distinct temporal thresholds, such as career ages beyond which Level 2 grants are no longer observed, or before which Level 1A status is rarely attained. Surprisingly, this clear stratification vanishes when examining the $h$-index, despite its prominence as a primary indicator of citation-based productivity. Instead, grant levels appear distributed without clear patterns along the $h$-index axis. Furthermore, significant variance in $h$-index values is observed across all age cohorts, which raises critical questions regarding the actual weight afforded to scientific impact versus the weight implicitly placed on seniority itself.

\begin{figure}[htbp]
    \centering
    \includegraphics[width=1.0\linewidth]{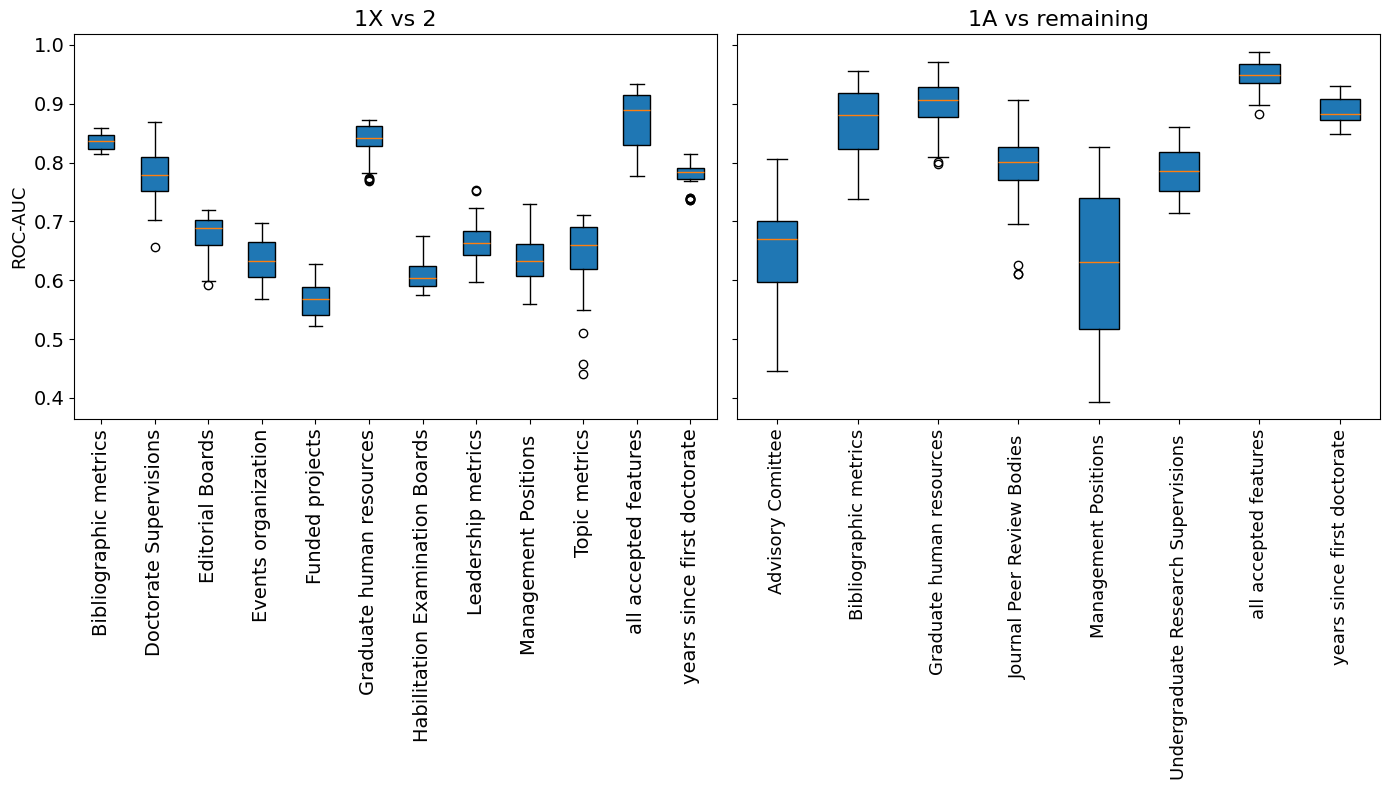}
    \caption{box plot of the ROC-AUC of the models for many combinations of features. }
    \label{fig:boxplot}
\end{figure}

Finally, we assess the independent contribution of the selected feature blocks. To do so, two proposed models (identical to those in Tables \ref{tab:auc_comparison1} and \ref{tab:auc-comparison2}) were trained using a 5-fold cross-validation process. Each model was trained using only a single feature block from those unanimously identified as relevant (as shown in Table \ref{tab:feature_freq_comparison}). Additionally, we included a step evaluating \emph{years since doctorate} in isolation to further examine its individual predictive potential, as well as a baseline step considering all selected features. The results are compiled in the box plots shown in Figure \ref{fig:boxplot}.
Most predictors, while carrying a meaningful signal, are weak when used individually and only form a robust model as a collective. Despite this, \emph{bibliographic metrics}, \emph{graduate human resources} and \emph{years since doctorate} each achieved performance significantly above average, nearly matching the composite performance. This suggests that these are the strongest individual predictors of grant level, although the remaining features still provide valuable incremental information.

In conclusion, it was possible to ascertain that in practice, the feature set can be successfully reduced into few, highly significant features, while many indicated features carry no additional signal. This suggests that, in practice, a reduced set of criteria would in practice be more informative of the actual process behind grant classifications. This also motivates investigation into the underappreciated attributes and their relationship with high performance in other areas.

\section{Conclusion}

This study empirically examined the practical validity of the official evaluation criteria used in the Research Productivity Grant (PQ) framework in Computer Science. By operationalizing regulation-defined dimensions as measurable features and subjecting them to a conservative, block-based feature selection strategy, we treated the stated criteria not as assumptions, but as testable hypotheses. The consistently high classification performance indicates that PQ levels carry a strong and structured statistical signal that can be recovered from structured curriculum and bibliometric data.

Our results reveal a clear concentration of explanatory power in a relatively small subset of dimensions. Bibliographic production, graduate human resources formation, and management-related roles emerged as the most robust and universally informative components across tasks. In contrast, several criteria explicitly emphasized in the regulations, particularly technological production, patents, software registration, instructional materials, and certain service activities, showed no detectable statistical contribution to classification outcomes. Importantly, removing these features did not degrade predictive performance, suggesting that, in practice, the effective evaluative signal is substantially more compact than the formal framework implies.

Differences between the two classification tasks further indicate that evaluation standards are not uniform across grant levels. The distinction between baseline productivity (1X vs 2) appears to incorporate scrutiny over leadership, topical focus, and funding capacity, whereas the identification of top-tier researchers (1A) is driven predominantly by core bibliometric and graduate supervision indicators, with awards playing a secondary role. This asymmetry suggests that certain policy-defined dimensions may exert influence only at specific decision thresholds.

Overall, the findings seem to point to the fact that, While the PQ system appears internally consistent and highly predictable from structured data, the effective criteria seem narrower than formally stated. These results contribute to ongoing discussions about transparency and accountability in research assessment, demonstrating how interpretable machine-learning methodologies can be employed to audit policy-defined evaluation systems and inform evidence-based refinement of productivity frameworks. In future work, we intend to incorporate additional data dimensions to enhance predictive performance. Specifically, we aim to employ complex network analysis~\citep{brito2020complex} to investigate how the topology of research collaboration networks might influence scientific success~\citep{coautoria}.

\section*{Acknowledgments}

The authors gratefully acknowledge financial support from the São Paulo Research Foundation (FAPESP) (grant nos. 2025/00944-6 and 2025/16076-3) and the National Council for Scientific and Technological Development (CNPq-Brazil) (grant no. 304189/2025-1).

\bibliographystyle{abbrvnat} 

\newpage

\section*{Supplementary Information}

The following supplementary material includes additional tables and figures that complement the results presented in the main manuscript

\renewcommand{\thetable}{S\arabic{table}}
\setcounter{table}{0}
\renewcommand{\thefigure}{S\arabic{figure}}
\setcounter{figure}{0}

\begin{table}[htbp]
\centering
\caption{Bibliographic production criteria and extracted features}
\label{Table1}
\small
\begin{tabularx}{\linewidth}{p{0.45\linewidth} p{0.55\linewidth}}
\toprule
\textbf{Mentioned metrics} & \textbf{Extracted features} \\
\midrule
Number of publications &
Number of journal articles published*; Number of conference papers published* \\

Quality of publication venues &
Number of publications on high-impact venues*; Number of publications on foreign venues* \\

Regularity and continuity of production &
Average amount of yearly publications in the period*; Number of lines of research \\

Researcher’s intellectual leadership &
Amount of publications as first author* \\

Structural limits &
Conference to journal publication ratio*; Number of publications not related to computer science*; Not related to computer science to related to computer science ratio* \\

Citations received &
Number of citations received in the period* \\

Impact indices &
Author h-index*; Author i10-index*; Number of field-normalized highly cited publications* \\

Avoidance of artificial co-authorship &
Average repeated co-authorship degree* \\

Rigor of the peer-review process &
-- \\
\bottomrule
\end{tabularx}
\end{table}

\begin{table}[htbp]
\centering
\caption{Human resources formation criteria and extracted features}
\label{Table2}
\small
\begin{tabularx}{\linewidth}{p{0.45\linewidth} p{0.55\linewidth}}
\toprule
\textbf{Mentioned metrics} & \textbf{Extracted features} \\
\midrule
Supervision of Master’s and PhD students &
Master’s thesis supervisions; Doctoral dissertation supervisions \\

Supervision of postdoctoral researchers &
Postdoctoral supervisions \\

Supervision of undergraduate research &
Undergraduate research supervisions \\

Course and teaching material development &
Special courses taught; Instructional material development \\

Participation in examination committees &
Participation in doctoral committees; Participation in master’s committees; Participation in habilitation committees \\

Research group creation and consolidation &
-- \\

Participation of students in research outcomes &
-- \\

Awards linked to supervised theses &
-- \\
\bottomrule
\end{tabularx}
\end{table}

\begin{table}[htbp]
\centering
\caption{National and international recognition criteria}
\label{Table3}
\small
\begin{tabularx}{\linewidth}{p{0.45\linewidth} p{0.55\linewidth}}
\toprule
\textbf{Mentioned metrics} & \textbf{Extracted features} \\
\midrule
National and international engagement &
Organization of national events; Organization of international events \\

Committee, editorial, and collaboration service &
Editorial board memberships; Peer reviewing for journals; Service on advisory committees; Commission board memberships \\

Awards and honors &
Awards and honors \\

Scientific leadership and management &
Management positions in scientific institutions \\

Evaluation and funding committees &
Peer reviewing for funding agencies \\

Public policy contributions &
-- \\
\bottomrule
\end{tabularx}
\end{table}

\begin{table}[htbp]
\centering
\caption{Technological, social, and governmental contributions}
\label{Table4}
\small
\setlength{\tabcolsep}{6pt}
\begin{tabularx}{\linewidth}{p{0.45\linewidth} p{0.55\linewidth}}
\toprule
\textbf{Mentioned metrics} & \textbf{Extracted features} \\
\midrule
Technology transfer &
Technologies developed in Brazil; Technologies developed abroad \\

Patents &
Patents registered in Brazil; Patents registered abroad \\

Registered software &
Software registered in Brazil; Software registered abroad \\

Company creation &
-- \\

Industrial adoption &
-- \\

Public policy impact &
-- \\

Demonstrated social impact &
-- \\
\bottomrule
\end{tabularx}
\end{table}

\begin{table}[htbp]
\centering
\caption{Participation and funding of research projects and networks}
\label{Table5}
\small
\begin{tabularx}{\linewidth}{p{0.45\linewidth} p{0.55\linewidth}}
\toprule
\textbf{Mentioned metrics} & \textbf{Extracted features} \\
\midrule
Participation in funded projects &
Funded research projects \\

Collaborative research networks &
Average amount of different authors collaborated with in a year*; 
Average amount of different institutions of authors collaborated with in a year*; 
Average amount of foreign authors collaborated with in a year* \\

Project roles &
-- \\
\bottomrule
\end{tabularx}
\end{table}

\begin{figure}[htbp]
    \centering
    \includegraphics[width=1.0\linewidth]{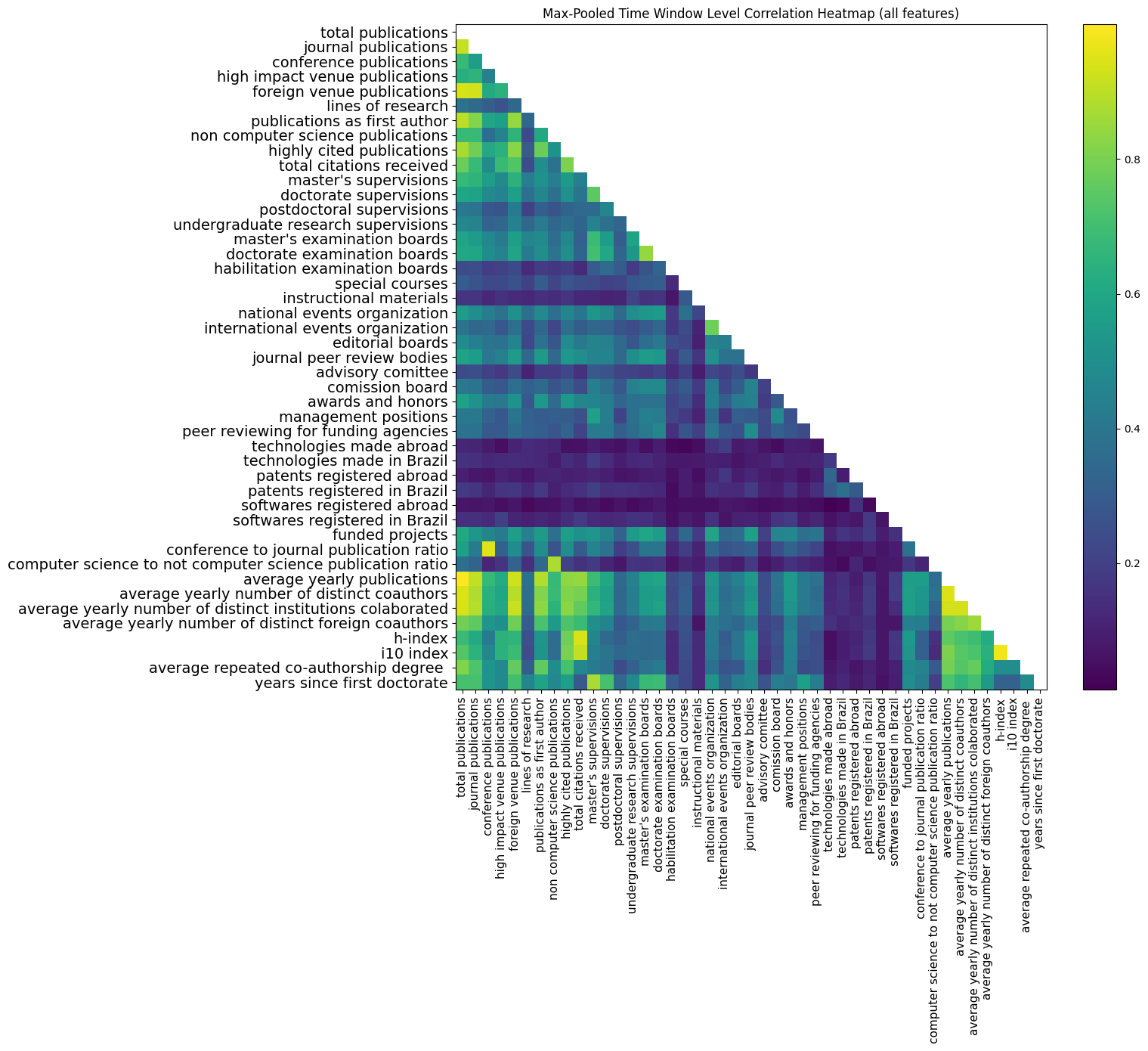}
    \caption{Heatmap of maximum absolute correlations between all features across time windows}
    \label{fig:heatmap_full}
\end{figure}

\end{document}